\documentclass[graybox,vecphys]{svmult}

\usepackage{mathptmx}       % selects Times Roman as basic font
\usepackage{helvet}         % selects Helvetica as sans-serif font
\usepackage{courier}        % selects Courier as typewriter font
\usepackage{type1cm}        % activate if the above 3 fonts are
\usepackage{makeidx}         % allows index generation
\usepackage{graphicx}        % standard LaTeX graphics tool
\usepackage{multicol}        % used for the two-column index
\usepackage[bottom]{footmisc}% places footnotes at page bottom
\usepackage{url}
\usepackage{slashed}
\usepackage{braket}
\usepackage{tensor}
\usepackage{combelow}

\makeindex             % used for the subject index
\begin{document}

\title*{Fermions on adS}
\author{Victor E. Ambru\cb{s} and Elizabeth Winstanley}
\institute{Victor E. Ambru\c{s} \at Consortium for Fundamental Physics, School of Mathematics and Statistics, University of Sheffield, Hicks Building, Hounsfield Road, Sheffield, S3 7RH, United Kingdom \\\email{app10vea@sheffield.ac.uk}
\and Elizabeth Winstanley \at Consortium for Fundamental Physics, School of Mathematics and Statistics, University of Sheffield, Hicks Building, Hounsfield Road, Sheffield, S3 7RH, United Kingdom \\\email{E.Winstanley@sheffield.ac.uk}}
\maketitle

% \abstract*{We construct the Feynman propagator for Dirac fermions on anti-de Sitter 
% space-time and present an analytic expression for the bi-spinor of parallel transport.
% We then renormalise the vacuum expectation value of the stress-energy tensor
% and end by analysing its renormalised expectation value at finite temperatures.}

\abstract{We construct the Feynman propagator for Dirac fermions on anti-de Sitter 
space-time and present an analytic expression for the bi-spinor of parallel transport.
We then renormalise the vacuum expectation value of the stress-energy tensor
and end by analysing its renormalised expectation value at finite temperatures.}

\section{Introduction}\label{sec:intro}

Quantum field theory (QFT) on curved spaces (CS) is a semi-classical theory for
the investigation of quantum effects in gravity. Due to its simplicity, the scalar field has been the main focus of 
QFT on CS. However, due to the fundamental difference 
between the quantum behaviour of fermions and bosons, it is important to also study fermionic fields.
In this paper, we consider the propagation of Dirac fermions
on the anti de Sitter space-time (adS) background space-time, where the maximal symmetry can be used to 
obtain analytic results.

We start this paper by presenting in Sec.~\ref{sec:ads} an expression for the spinor parallel propagator \cite{art:muck}.
Using results from geodesic theory \cite{art:allen_jacobson,art:muck}, an exact expression for the Feynman propagator is obtained 
in Sec.~\ref{sec:sf}. Section \ref{sec:had} is devoted to Hadamard's regularisation method \cite{art:najmi_ottewill}, while, 
in Sec.~\ref{sec:tvac}, the
result for the renormalised vacuum expectation value (v.e.v.) of the stress-energy tensor (SET) is presented using two methods:
the Schwinger-de Witt method \cite{art:christensen} and the Hadamard method \cite{art:hack}. The exact form of the bi-spinor of
parallel transport is then used in Sec.~\ref{sec:tbeta} to calculate the thermal expectation value (t.e.v.) of the SET 
for massless spinors. More details on the current work, as well as an extension to massive spinors, can be found in 
\cite{art:ambrus_winstanley_ads}.

\section{Geometric structure of adS}\label{sec:ads}
Anti-de Sitter space-time (adS) is a vacuum solution of the Einstein equation
with a negative cosmological constant, having the following line element:
\begin{equation}\label{eq:ds2}
 ds^2 = \frac{1}{\cos^2\omega r}
 \left[-dt^2 + dr^2 + \frac{\sin^2\omega r}{\omega^2} \left(d\theta^2 + \sin^2\theta d\varphi^2\right)\right].
\end{equation}
The time coordinate $t$ runs from $-\infty$ to $\infty$, thereby giving the covering space of adS. The radial
coordinate $r$ runs from $0$ to the space-like boundary at $\pi/2\omega$, while $\theta$ and $\varphi$
are the usual elevation and azimuthal angular coordinates. In the Cartesian gauge, the line element (\ref{eq:ds2}) 
admits the following natural frame \cite{art:cota}:
\begin{equation}\label{eq:tetrad}
 \omega^{\hat{t}} = \frac{\D t}{\cos\omega r}, \qquad
 \omega^{\hat{i}} = \frac{\D x^j}{\cos\omega r}\left[\frac{\sin\omega r}{\omega r}\left(\delta_{ij} - \frac{x^ix^j}{r^2}\right) + 
 \frac{x^ix^j}{r^2}\right],
\end{equation}
such that $\eta_{\hat{\alpha}\hat{\beta}} \omega^{\hat{\alpha}}_\mu \omega^{\hat{\beta}}_\nu = g_{\mu\nu}$, where
$\eta_{\hat{\alpha}\hat{\beta}} = \rm{diag}(-1,1,1,1)$ is the Minkowski metric.

A key role in the construction of the propagator of the Dirac field is played by the bi-spinor of parallel transport
$\Lambda(x,x')$, which satisfies the parallel transport equation $n^\mu D_\mu \Lambda(x,x') = 0$ \cite{art:muck}. On adS, 
the explicit form of $\Lambda(x,x')$ is \cite{art:ambrus_winstanley_ads}: 
\begin{eqnarray}
 \Lambda(x,x') &=& \frac{\cos(\omega\Delta t/2)}{\cos(\omega s/2)\sqrt{\cos\omega r \cos \omega r'}}
 \Big\{
 \cos\frac{\omega r}{2} \cos\frac{\omega r'}{2} + 
 \frac{\mathbf{x}\cdot \hat{\mathbf{\gamma}}}{r} \frac{\mathbf{x'}\cdot \hat{\mathbf{\gamma}}}{r'}
 \sin\frac{\omega r}{2} \sin\frac{\omega r'}{2}\nonumber\\
 && - \gamma^{\hat{t}} \tan\frac{\omega\Delta t}{2} \left(
 \frac{\mathbf{x}\cdot \hat{\mathbf{\gamma}}}{r} \sin\frac{\omega r}{2} \cos\frac{\omega r'}{2} - 
 \frac{\mathbf{x'}\cdot \hat{\mathbf{\gamma}}}{r'} \cos\frac{\omega r}{2} \sin\frac{\omega r'}{2}\right)\Big\},
 \label{eq:lambda}
\end{eqnarray}
where $\gamma^{\hat{\alpha}} = (\gamma^{\hat{t}}, \hat{\mathbf{\gamma}})$ are the gamma matrices in the Dirac representation
and $s$ is the geodesic distance between $x$ and $x'$.
% The parallel transport of vectors and tensors is performed in a similar fashion by the bi-vector of parallel transport
% $g_{\mu\nu'}(x,x')$, which satisfies the equation $n^\lambda \nabla_\lambda g_{\mu\nu'} = 0$ \cite{art:allen_jacobson}. 
% An analytic expression can be obtained for $g_{\mu\nu'}$ as well, however, its exact form must be omitted due to 
% lack of space.

\section{Feynman propagator on adS}\label{sec:sf}
The Feynman propagator $S_F(x,x')$ for a Dirac field of mass $m$ can be defined as the solution of the inhomogeneous 
Dirac equation, with appropriate boundary conditions:
\begin{equation}\label{eq:sf_def}
 (\I\slashed{D} - m) S_F(x,x') = (-g)^{-1/2} \delta^4(x-x'),
\end{equation}
where $D_\mu$ denotes the spinor covariant derivative and $g$ is the determinant of the background space-time metric.
Due to the maximal symmetry of adS, the Feynman propagator can be written in the following form \cite{art:muck}:
\begin{equation}\label{eq:sf_muck}
 S_F(x,x') = \left[\alpha_F(s) + \slashed{n}\, \beta_F(s)\right]\, \Lambda(x,x').
\end{equation}
The functions $\alpha_F$ and $\beta_F$ can be determined using (\ref{eq:sf_def}):
\begin{eqnarray}
 \alpha_F &=& \frac{\omega^3 k}{16\pi^2} \cos\frac{\omega s}{2}\left\{
 -\frac{1}{\sin^2\frac{\omega s}{2}}
 + 2 (k^2 - 1) \ln \left|\sin\frac{\omega s}{2}\right| {}_2F_1\left(2+k,2-k;2;\sin^2\frac{\omega s}{2}\right)
 \right.\nonumber\\
 & &\left. + (k^2 - 1) \sum_{n=0}^\infty \frac{(2+k)_n (2-k)_n}{(2)_n n!} \left(\sin^2\frac{\omega s}{2}\right)^n
 \Psi_n\right\},\label{eq:alpha}\label{eq:af}\\
 \beta_F &=& \frac{\I \omega^3}{16\pi^2}  \sin\frac{\omega s}{2}
 \Bigg\{\nonumber\\
 & & \frac{1 + k^2 \sin^2(\omega s/2)}{[\sin(\omega s/2)]^4}
 - k^2 (k^2 - 1) \ln \left|\sin\frac{\omega s}{2}\right| 
 {}_2F_1\left(2+k,2-k;3;\sin^2\frac{\omega s}{2}\right) \nonumber\\
 & &\left. - \frac{k^2 (k^2 - 1)}{2} \sum_{n=0}^\infty \frac{(2+k)_n (2-k)_n}{(3)_n n!} \left(\sin^2\frac{\omega s}{2}\right)^n
 \left(\Psi_n - \frac{1}{2+n}\right)\right\}\label{eq:beta},\label{eq:bf}
\end{eqnarray}
where $a_n = \Gamma(a+n)/\Gamma(a)$ is the Pochhammer symbol,
$\Gamma(z) = \int_0^\infty x^{z-1} \E^{-x} dx$ is the gamma function,
$k = m/\omega$, 
\begin{equation}
 \Psi_n = \psi(k + n + 2) + \psi(k - n - 1) - \psi(n + 2) - \psi(n + 1)
\end{equation}
and $\psi(z) = \D \ln \Gamma(z) / \D z$ is the digamma function.

\section{Hadamard renormalisation}\label{sec:had}
To regularise $S_F$, it is convenient to use the auxilliary propagator $\mathcal{G}_F$, defined by analogy to 
flat space-time \cite{art:najmi_ottewill}:
\begin{equation}
 S_F(x,x') = (\I \slashed{D} + m) \mathcal{G}_F.
\end{equation}
On adS, $\mathcal{G}_F$ can be written using the bi-spinor of parallel transport:
\begin{equation}
 \mathcal{G}_F(x,x') = \frac{\alpha_F}{m} \Lambda(x,x'),
\end{equation}
where $\alpha_F$ is given in (\ref{eq:af}).

According to Hadamard's theorem, the divergent part $\mathcal{G}_H$ of $\mathcal{G}_F$ is state-independent, 
having the form \cite{art:najmi_ottewill}:
\begin{equation}
 \mathcal{G}_H(x,x') = \frac{1}{8\pi^2} \left[\frac{u(x,x')}{\sigma} + v(x,x') \ln \mu^2 \sigma\right],
\end{equation}
where $u(x,x')$ and $v(x,x')$ are finite when $x'$ approaches $x$, $\sigma = -s^2/2$ is Synge's world function and
$\mu$ is an arbitrary mass scale.
The functions $u$ and $v$ can be found by solving the inhomogeneous Dirac equation (\ref{eq:sf_def}), requiring that 
the regularised auxilliary propagator $\mathcal{G}_F^{\rm{reg}} \equiv \mathcal{G}_F - \mathcal{G}_H$ is finite
in the coincidence limit:
\begin{eqnarray}
 u(x,x') &=& \sqrt{\Delta(x,x')} \Lambda(x,x'),\label{eq:uhad}\\
 v(x,x') &=& \frac{\omega^2}{2} (k^2 - 1) \cos\frac{\omega s}{2} {}_2F_1\left(2-k,2+k;2;\sin^2\frac{\omega s}{2}\right) \Lambda(x,x'),
 \label{eq:vhad}
\end{eqnarray}
where the Van Vleck-Morette determinant $\Delta(x,x') = (\omega s / \sin \omega s)^3$ on adS.
%%%%%%%%%%%%%%%%%%%%%%%%%%%%%%%%%%%%%%%%%%%%%%%%%%%%%%%%%%%%%%%%%%%%%%%%%%%%%%% 
\section{Renormalised vacuum stress-energy tensor}\label{sec:tvac}
To remove the traditional divergences of quantum field theory, we employ two regularisation methods: the
Schwinger--de Witt method in Sec.~\ref{sec:tvac:sdw} and the Hadamard method in Sec.~\ref{sec:tvac:had}.
Due to the symmetries of adS, the regularised v.e.v. of the SET takes the form 
$\braket{\tens{T}_{\mu\nu}}_{\rm{vac}}^{\rm{reg}} = \frac{1}{4} \tens{T} g_{\mu\nu}$, 
where $\tens{T} = \tens{T}\indices{^\mu_\mu}$ is its trace. 
The renormalisation process has the profound consequence of shifting $\tens{T}$ 
for the massless (hence, conformal) Dirac field to a finite value, referred to as the conformal anomaly.
%%%%%%%%%%%%%%%%%%%%%%%%%%%%%%%%%%%%%%%%%%%%%%%%%%%%%%%%%%%%%%%%%%%%%%%%%%%%%%%
\subsection{Schwinger--de Witt regularisation}\label{sec:tvac:sdw}
By using the Schwinger--de Witt approach to investigate the singularity structure of the propagator of the
Dirac field in the coincidence limit, Christensen \cite{art:christensen} calculates a set of subtraction terms
which only depend on the geometry of the background space-time, using the following formula:
\begin{equation}\label{eq:tmunu_sf_can}
 \braket{\tens{T}_{\mu\nu}} = \lim_{x'\rightarrow x} {\rm{tr}}
 \left\{\frac{\I}{2} \left[\gamma_{(\mu} D_{\nu)} - \gamma_{(\mu'}D_{\nu')}\right] S_F(x,x')\right\}.
\end{equation}
After subtracting Christensen's terms, we exactly recover the result obtained by Camporesi and 
Higuchi \cite{art:camporesi_higuchi} using the Pauli-Villars regularisation method:
\begin{equation}\label{eq:tvac:sdw}
 \braket{\tens{T}}_{\rm{vac}}^{\rm{SdW}} = -\frac{\omega^4}{4\pi^2} \left\{
 \frac{11}{60} + k - \frac{k^2}{6} - k^3 +
 2 k^2(k^2 -1) \left[\ln\frac{\mu}{\omega} - \psi(k)\right]\right\},
\end{equation}
where $\mu$ is an arbitrary mass scale.
%%%%%%%%%%%%%%%%%%%%%%%%%%%%%%%%%%%%%%%%%%%%%%%%%%%%%%%%%%%%%%%%%%%%%%%%%%%%%%%
\subsection{Hadamard regularisation}\label{sec:tvac:had}
The Hadamard theorem presented in Sec.~\ref{sec:had} allows the renormalisation to be performed at the level of the 
propagator. To preserve the conservation of the SET, the following definition for the SET must be used \cite{art:hack}:
\begin{equation}\label{eq:tmunu_sf_hack}
 \braket{\tens{T}_{\mu\nu}} = \lim_{x'\rightarrow x} {\rm{tr}}
 \left\{\frac{\I}{2} \left[\gamma_{(\mu} D_{\nu)} - \gamma_{(\mu'}D_{\nu')}\right] + \frac{1}{6} g_{\mu\nu}
 \left[\frac{\I}{2} (\slashed{D} - \slashed{D}') - m\right] \right\} S^{\rm{reg}}_F(x,x'),
\end{equation}
where $S_F^{\rm{reg}}(x,x') = (\I \slashed{D} + m) (\mathcal{G}_F - \mathcal{G}_H)$ is the regularised propagator.
The coefficient of $g_{\mu\nu}$ is proportional to the Lagrangian of the Dirac field and evaluates to zero
when applied to a solution of (\ref{eq:sf_def}). However, $S_F^{\rm{reg}}(x,x')$ is not a solution of 
(\ref{eq:sf_def}). The v.e.v. obtained from (\ref{eq:tmunu_sf_hack}) matches perfectly the result 
obtained by Camporesi and Higuchi \cite{art:camporesi_higuchi} using the zeta-function regularisation 
method ($\gamma$ is Euler's constant):
\begin{equation}\label{eq:tvac:had}
 \braket{\tens{T}}_{\rm{vac}}^{\rm{Had}} = -\frac{\omega^4}{4\pi^2}\left\{
 \frac{11}{60} + k - \frac{7k^2}{6} - k^3 + \frac{3k^4}{2} + 
 2k^2(k^2 - 1)\left[\ln \frac{\mu \E^{-\gamma} \sqrt{2}}{\omega} - \psi(k)\right]\right\}.
\end{equation}
Even though the results (\ref{eq:tvac:sdw}) and (\ref{eq:tvac:had}) are different for general values
of the mass parameter $k$, they yield the same conformal anomaly.
We would like to stress that the omission of the term proportional to $g_{\mu\nu}$ in (\ref{eq:tmunu_sf_hack})
would increase the value of the conformal anomaly by a factor of $3$.
%%%%%%%%%%%%%%%%%%%%%%%%%%%%%%%%%%%%%%%%%%%%%%%%%%%%%%%%%%%%%%%%%%%%%%%%%%%%%%%
\section{Thermal stress-energy tensor}\label{sec:tbeta}
\begin{figure}
\begin{tabular}{cc}
 \includegraphics[width=0.45\linewidth]{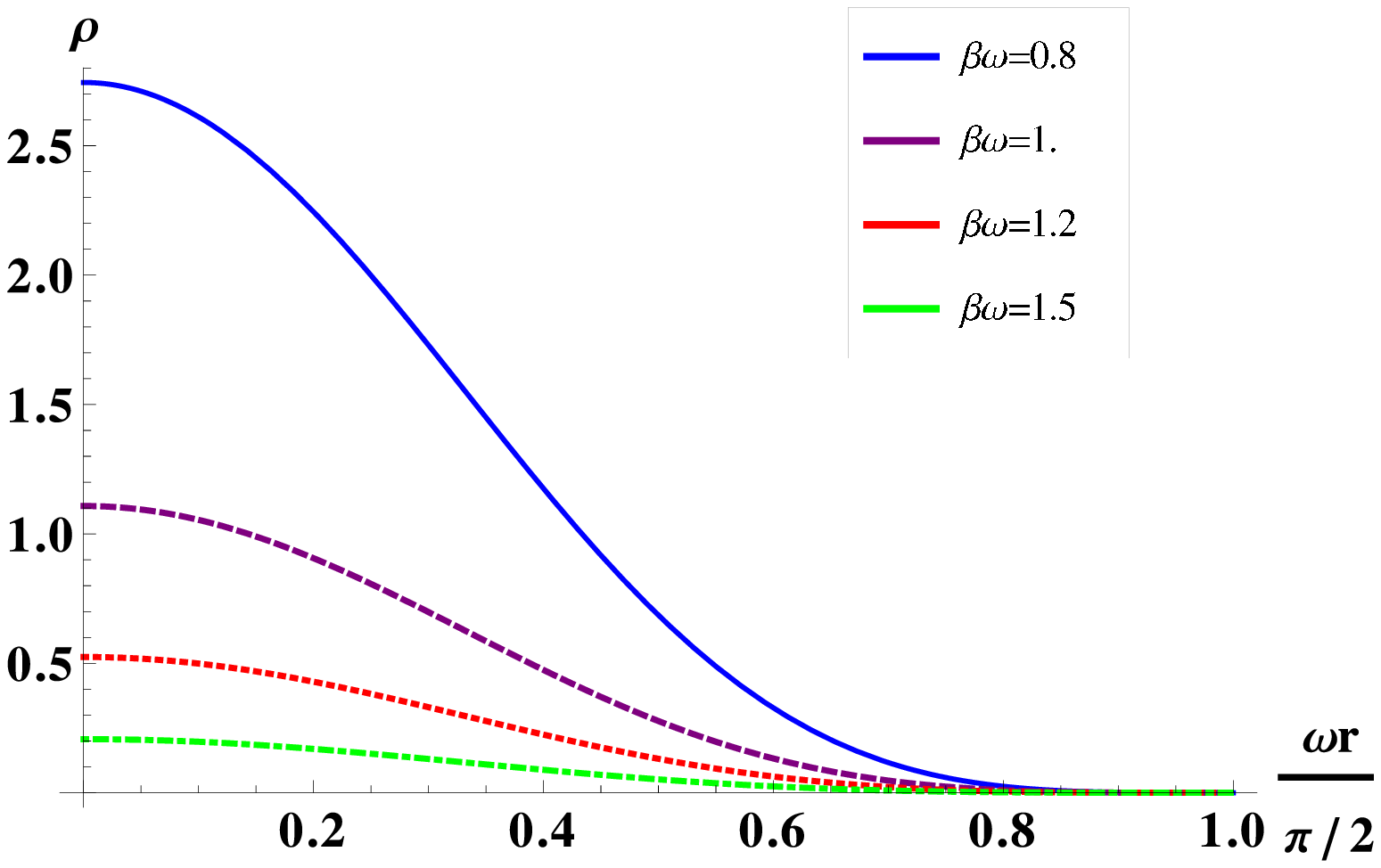} &
 \includegraphics[width=0.45\linewidth]{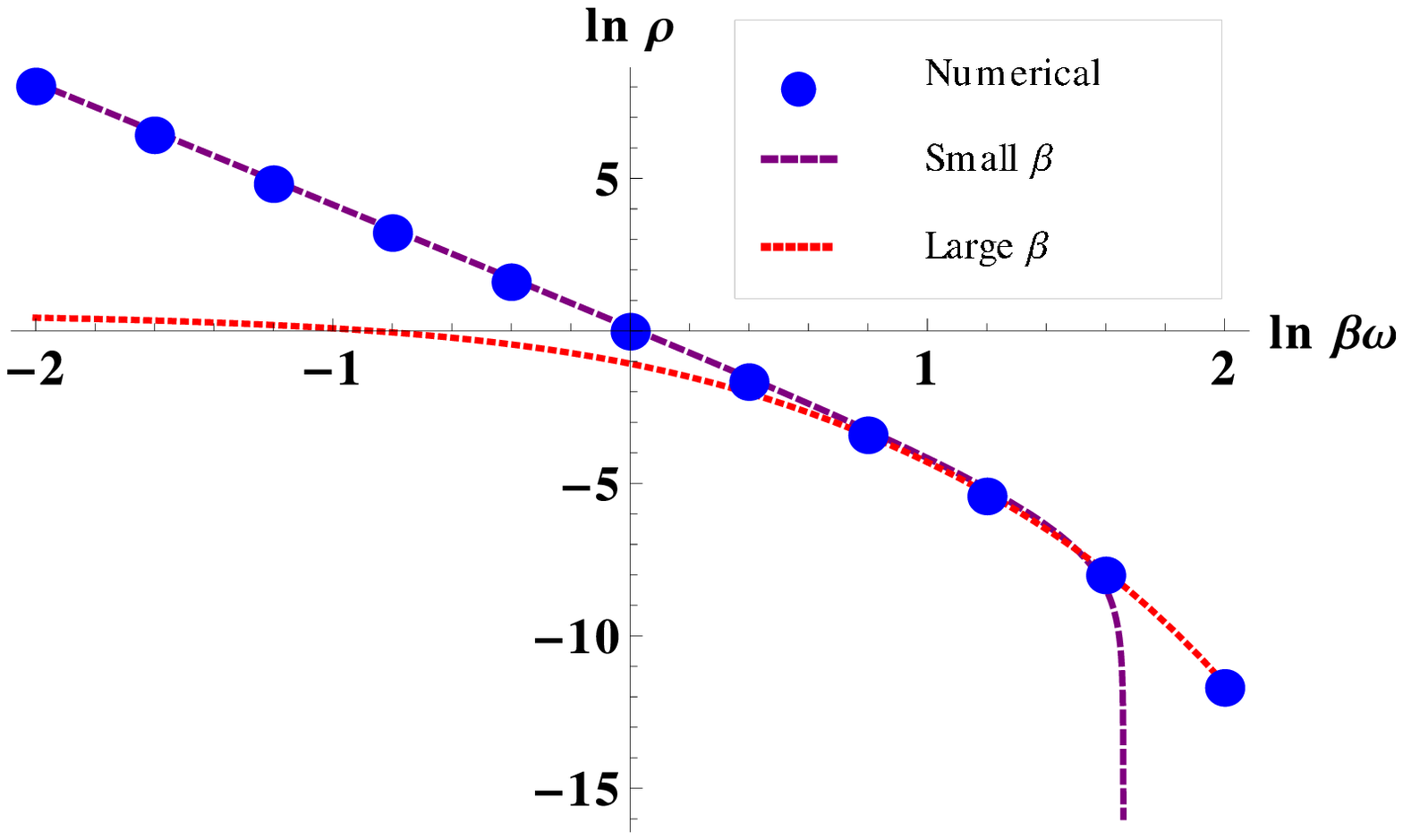}
\end{tabular}
\caption{{\bf a}\, $\varrho$ between the origin ($r=0$) and the boundary ($r\omega = \pi/2$) for $\beta \omega = 0.8$,$1.0$,$1.2$ and $1.4$; 
{\bf b}\, Log-log plot of $\varrho$ in terms of $\beta \omega$; comparison with the asymptotic results in (\ref{eq:rho-small}) and (\ref{eq:rho-large})}
\label{fig}
\end{figure}

The renormalised thermal expectation value (t.e.v.) of the SET can be written as:
\begin{equation}
 \braket{\tens{T}_{\mu\nu}}_\beta^{\rm{reg}} = \braket{:\tens{T}_{\mu\nu}:}_\beta + \braket{\tens{T}_{\mu\nu}}_{\rm{vac}}^{\rm{ren}},
\end{equation}
where $\beta = T^{-1}$ is the inverse temperature and the colons $::$ indicate that the operator enclosed is in normal
order, i.e. with its v.e.v. subtracted. The bi-spinor of parallel transport can be used to show that
\begin{equation}
 \braket{:\tens{T}^\mu_{\phantom{\mu}\nu}:}_\beta = \rm{diag}(-\varrho, p, p, p),
\end{equation}
where $\rho$ is the energy density and $p$ is the pressure. If $m = 0$, we have $p = \rho / 3$ and:
\begin{equation}
 \varrho\rfloor_{m = 0} = -\frac{3 \omega^4}{4\pi^2} (\cos\omega r)^4 \sum_{j = 1}^\infty (-1)^j
 \frac{\cosh(j \omega \beta/2)}{[\sinh (j \omega \beta / 2)]^4},
\end{equation}
with the coordinate dependence fully contained in the $(\cos \omega r)^4$ prefactor. 
The first term in the sum over $j$ is within $6\%$ of the sum, while the first two terms together are less
than $1\%$ away, for all values of $\omega \beta$. The small and large $\omega \beta$ limits can be extracted:
\begin{eqnarray}
 \varrho\rfloor_{m = 0} &=& (\cos \omega r)^4 \left[\frac{7\pi^2}{60\beta^4} - \frac{\omega^2}{24\beta^2} + O(\omega^4)\right],\label{eq:rho-small}\\
 \varrho\rfloor_{m = 0} &=& \frac{6\omega^4}{\pi^2} \frac{(\cos\omega r)^4}{1 + \E^{3 \beta \omega / 2}} 
 \left[1 + 5 \E^{-\omega \beta} \frac{1 + \E^{-3 \omega \beta / 2}}{1 + \E^{-5 \omega \beta / 2}} +
 O(\E^{-2\omega \beta})\right].\label{eq:rho-large}
\end{eqnarray}
Figure~\ref{fig} shows a graphical representation of the above results.

\begin{acknowledgement}
This work is supported by the Lancaster-Manchester-Sheffield Consortium for 
Fundamental Physics under STFC grant ST/J000418/1,
the School of Mathematics and Statistics at the University of Sheffield 
and European Cooperation in Science and Technology (COST) action MP0905 
``Black Holes in a Violent Universe''.
\end{acknowledgement}

%%%%%%%%%%%%%%%%%%%%%%%% referenc.tex %%%%%%%%%%%%%%%%%%%%%%%%%%%%%%
% sample references
% %
% Use this file as a template for your own input.
%
%%%%%%%%%%%%%%%%%%%%%%%% Springer-Verlag %%%%%%%%%%%%%%%%%%%%%%%%%%
%
% BibTeX users please use
% \bibliographystyle{}
% \bibliography{}

\begin{thebibliography}{99.}%
% Journal article
\bibitem{art:allen_jacobson} Allen, B. and Jacobson, T.: Vector two-point functions in maximally symmetric spaces. 
Commun. Math. Phys. \textbf{103}, 669--692 (1986) 

\bibitem{art:ambrus_winstanley_ads} Ambru\cb{s}, V. E. and Winstanley, E.: Fermions on adS. 
\textit{Paper in preparation.} 

\bibitem{art:camporesi_higuchi} Camporesi, R. and Higuchi, A.: Stress-energy tensors in anti-de Sitter spacetime.
Phys. Rev. D \textbf{45}, 3591--3603 (1992)

\bibitem{art:christensen} Christensen, S. M.: Regularization, renormalization, and covariant geodesic point separation.
Phys. Rev. D \textbf{17}, 946--963 (1978)

\bibitem{art:cota} Cot\u{a}escu, I.: Dirac fermions in de Sitter and anti-de Sitter backgrounds. 
Rom. J. Phys. \textbf{52}, 895--940 (2007)

\bibitem{art:hack} Dappiaggi, C.; Hack, T.-P. and Pinamonti, N.: The extended algebra of observables for Dirac fields and the trace anomaly of their stress-energy tensor.
Rev. Math. Phys \textbf{21}, 1241--1312 (2009)

\bibitem{art:muck} M\"uck, W.: Spinor parallel propagator and Green's function in maximally symmetric spaces. 
J. Phys. A \textbf{33}, 3021--3026 (2000)

\bibitem{art:najmi_ottewill} Najmi, A.-H. and Ottewill, A. C.: Quantum states and the Hadamard form. II. Energy minimization for spin-1/2 fields. 
Phys. Rev. D \textbf{30}, 2573--2578 (1984)
% Use the following syntax and markup for your references if 
% the subject of your book is from the field 
% "Mathematics, Physics, Statistics, Computer Science"
%
% Contribution 
%\bibitem{science-contrib} Broy, M.: Software engineering --- from auxiliary to key technologies. In: Broy, M., Dener, E. (eds.) Software Pioneers, pp. 10-13. Springer, Heidelberg (2002)
%
% Online Document
%\bibitem{science-online} Dod, J.: Effective substances. In: The Dictionary of Substances and Their Effects. Royal Society of Chemistry (1999) Available via DIALOG. \\
%\url{http://www.rsc.org/dose/title of subordinate document. Cited 15 Jan 1999}
%
% Monograph
%\bibitem{science-mono} Geddes, K.O., Czapor, S.R., Labahn, G.: Algorithms for Computer Algebra. Kluwer, Boston (1992) 
%
%
% Journal article by DOI
%\bibitem{science-DOI} Slifka, M.K., Whitton, J.L.: Clinical implications of dysregulated cytokine production. J. Mol. Med. (2000) doi: 10.1007/s001090000086 
%
%\bigskip

% Use the following (APS) syntax and markup for your references if 
% the subject of your book is from the field 
% "Mathematics, Physics, Statistics, Computer Science"
%
% Online Document
% \bibitem{phys-online} J. Dod, in \textit{The Dictionary of Substances and Their Effects}, Royal Society of Chemistry. (Available via DIALOG, 1999), 
% \url{http://www.rsc.org/dose/title of subordinate document. Cited 15 Jan 1999}
%
% Monograph
% \bibitem{phys-mono} H. Ibach, H. L\"uth, \textit{Solid-State Physics}, 2nd edn. (Springer, New York, 1996), pp. 45-56 
% %
% % Journal article
% \bibitem{phys-journal} S. Preuss, A. Demchuk Jr., M. Stuke, Appl. Phys. A \textbf{61}
% %
% % Journal article by DOI
% \bibitem{phys-DOI} M.K. Slifka, J.L. Whitton, J. Mol. Med., doi: 10.1007/s001090000086
% %
% % Contribution 
% \bibitem{phys-contrib} S.E. Smith, in \textit{Neuromuscular Junction}, ed. by E. Zaimis. Handbook of Experimental Pharmacology, vol 42 (Springer, Heidelberg, 1976), p. 593
% %
% \bigskip
% %
% % Use the following syntax and markup for your references if 
% % the subject of your book is from the field 
% % "Psychology, Social Sciences"
% %
% %
% % Monograph
% \bibitem{psysoc-mono} Calfee, R.~C., \& Valencia, R.~R. (1991). \textit{APA guide to preparing manuscripts for journal publication.} Washington, DC: American Psychological Association.
% %
% % Online Document
% \bibitem{psysoc-online} Dod, J. (1999). Effective substances. In: The dictionary of substances and their effects. Royal Society of Chemistry. Available via DIALOG. \\
% \url{http://www.rsc.org/dose/Effective substances.} Cited 15 Jan 1999.
% %
% % Journal article
% \bibitem{psysoc-journal} Harris, M., Karper, E., Stacks, G., Hoffman, D., DeNiro, R., Cruz, P., et al. (2001). Writing labs and the Hollywood connection. \textit{J Film} Writing, 44(3), 213--245.
% %
% % Contribution 
% \bibitem{psysoc-contrib} O'Neil, J.~M., \& Egan, J. (1992). Men's and women's gender role journeys: Metaphor for healing, transition, and transformation. In B.~R. Wainrig (Ed.), \textit{Gender issues across the life cycle} (pp. 107--123). New York: Springer.
% %
% % Journal article by DOI
% \bibitem{psysoc-DOI}Kreger, M., Brindis, C.D., Manuel, D.M., Sassoubre, L. (2007). Lessons learned in systems change initiatives: benchmarks and indicators. \textit{American Journal of Community Psychology}, doi: 10.1007/s10464-007-9108-14.
% %
% %
% % Use the following syntax and markup for your references if 
% % the subject of your book is from the field 
% % "Humanities, Linguistics, Philosophy"
% %
% \bigskip
% %
% % Journal article
% \bibitem{humlinphil-journal} Alber John, Daniel C. O'Connell, and Sabine Kowal. 2002. Personal perspective in TV interviews. \textit{Pragmatics} 12:257--271
% %
% % Contribution 
% \bibitem{humlinphil-contrib} Cameron, Deborah. 1997. Theoretical debates in feminist linguistics: Questions of sex and gender. In \textit{Gender and discourse}, ed. Ruth Wodak, 99--119. London: Sage Publications.
% %
% % Monograph
% \bibitem{humlinphil-mono} Cameron, Deborah. 1985. \textit{Feminism and linguistic theory.} New York: St. Martin's Press.
% %
% % Online Document
% \bibitem{humlinphil-online} Dod, Jake. 1999. Effective substances. In: The dictionary of substances and their effects. Royal Society of Chemistry. Available via DIALOG. \\
% http://www.rsc.org/dose/title of subordinate document. Cited 15 Jan 1999
% %
% % Journal article by DOI
% \bibitem{humlinphil-DOI} Suleiman, Camelia, Daniel C. O�Connell, and Sabine Kowal. 2002. `If you and I, if we, in this later day, lose that sacred fire...�': Perspective in political interviews. \textit{Journal of Psycholinguistic Research}. doi: 10.1023/A:1015592129296.
% %
% %
% %
% \bigskip
% %
% %
% % Use the following syntax and markup for your references if 
% % the subject of your book is from the field 
% % "Computer Science, Economics, Engineering, Geosciences, Life Sciences"
% %
% %
% % Contribution 
% \bibitem{basic-contrib} Brown B, Aaron M (2001) The politics of nature. In: Smith J (ed) The rise of modern genomics, 3rd edn. Wiley, New York 
% %
% % Online Document
% \bibitem{basic-online} Dod J (1999) Effective Substances. In: The dictionary of substances and their effects. Royal Society of Chemistry. Available via DIALOG. \\
% \url{http://www.rsc.org/dose/title of subordinate document. Cited 15 Jan 1999}
% %
% % Journal article by DOI
% \bibitem{basic-DOI} Slifka MK, Whitton JL (2000) Clinical implications of dysregulated cytokine production. J Mol Med, doi: 10.1007/s001090000086
% %
% % Journal article
% \bibitem{basic-journal} Smith J, Jones M Jr, Houghton L et al (1999) Future of health insurance. N Engl J Med 965:325--329
% %
% % Monograph
% \bibitem{basic-mono} South J, Blass B (2001) The future of modern genomics. Blackwell, London 
% %
\end{thebibliography}
%

\end{document}